\documentclass[conference]{IEEEtran}

\usepackage[utf8]{inputenc} 
\usepackage[T1]{fontenc}

\usepackage{footmisc}

\usepackage[pdftex]{graphicx}

\usepackage[tight,footnotesize]{subfigure}

\usepackage{cite}

\usepackage[cmex10]{amsmath}
\usepackage{amssymb}
\usepackage{mathrsfs}
\usepackage{siunitx}

\usepackage{url}
\usepackage{lipsum}
\usepackage{multirow}
\usepackage{array}

\newcolumntype{L}[1]{>{\raggedright\let\newline\\\arraybackslash\hspace{0pt}}m{#1}}
\newcolumntype{C}[1]{>{\centering\let\newline\\\arraybackslash\hspace{0pt}}m{#1}}
\newcolumntype{R}[1]{>{\raggedleft\let\newline\\\arraybackslash\hspace{0pt}}m{#1}}

\newcommand{\vflow}{v_f}
\newcommand{\ntxk}[1]{N^{\text{Tx}}[#1]}
\newcommand{\nrxk}[2]{N^{\text{Rx}}_{#2}[#1]}
\newcommand{\nresidual}[1]{N_{\text{residual}}[#1]}
\newcommand{\isiwin}{m}
\newcommand{\symdur}{t_s}

\newcommand{\intdiff}{\Delta_c}

\newcommand{\bitones}{${\mbox{bit-1s}}$}
\newcommand{\bitone}{${\mbox{bit-1}}$}

\newcommand{\symbolTx}[1]{S [#1]}
\newcommand{\symbolRx}[1]{\hat{S} [#1]}

\newcommand{\shistorykm}[2]{\boldsymbol{H}_{#1}^{#2}}

\newcommand{\cpahistory}[1]{\boldsymbol{C}_{#1}}
\newcommand{\cpantx}[2]{n^{\cpahistory{#1}}_{#2}}

\newcommand{\diCO}{D}

\newcommand{\deltat}{\Delta{t}}
\newcommand{\deltaX}{\Delta{X}}
\newcommand{\deltaY}{\Delta{Y}}
\newcommand{\deltaZ}{\Delta{Z}}
\newcommand{\deltaXsub}[1]{\deltaX_{\text{#1}}}

\newcommand{\yakinsama}{\mathscr{N}\text{(}\mu\text{, }\sigma^{\text{2}}\text{)}}

\newcommand{\pk}[1]{p_{#1}}

\newcommand{\ensayisi}[1]{n_{#1}}

\newcommand{\prob}{\mathbf{P}}
\newcommand{\proberr}{\mathbf{P}_e}
\newcommand{\proberrsub}[1]{\mathbf{P}_{e|#1}}

\newcommand{\mystd}[1]{\text{STD}(#1)}

\newcommand{\nth}[1]{ \textit{$#1$}^{th}}
\newcommand{\rarrow}{\overrightarrow{r}}

\usepackage{color}
\usepackage{todonotes}
\definecolor{morange}{rgb}{0.8,0.2,0}
\definecolor{mblue}{rgb}{0,0.3,1.0}
\definecolor{mbluee}{rgb}{0.4,0.1,0.9}

\begin{document}
\title{MOL-Eye: A New Metric for the Performance Evaluation of a Molecular Signal}

\author{\IEEEauthorblockN{Meriç Turan$^{1}$, Mehmet Şükrü Kuran$^{2}$, H. Birkan Yilmaz$^{3}$, Chan-Byoung Chae$^{3}$, Tuna Tugcu$^{1}$}
\IEEEauthorblockA{$^{1}$Department of Computer Engineering, NETLAB, Bogazici University, Istanbul, Turkey\\
$^{2}$Department of Computer Engineering, Abdullah Gul University, Kayseri, Turkey\\
$^{3}$School of Integrated Technology, Yonsei Institute of Convergence Technology, Yonsei University, Seoul, South Korea\\
{E-mails:\{meric.turan, tugcu\}}@boun.edu.tr, sukru.kuran@agu.edu.tr, \{birkan.yilmaz, cbchae\}@yonsei.ac.kr
}
}


\maketitle

\begin{abstract}
Inspired by the eye diagram in classical radio frequency (RF) based communications, the \textit{MOL-Eye diagram} is proposed for the performance evaluation of a molecular signal within the context of molecular communication. Utilizing various features of this diagram, three new metrics for the performance evaluation of a molecular signal, namely the maximum eye height, standard deviation of received molecules, and counting SNR (CSNR) are introduced. The applicability of these performance metrics in this domain is verified by comparing the performance of binary concentration shift keying (BCSK) and BCSK with consecutive power adjustment (BCSK-CPA) modulation techniques in a vessel-like environment with laminar flow. The results show that, in addition to classical performance metrics such as bit-error rate and channel capacity, these performance metrics can also be used to show the advantage of an efficient modulation technique over a simpler one.

\end{abstract}


\IEEEpeerreviewmaketitle

\section{Introduction}
Nanonetworking is a communication paradigm that focuses on communication between nano-scale devices whose sizes are comparable to biological cells. Due to their small sizes, medical applications in in-vivo environments are expected to be one of the most prominent and driving application domains for these devices. However, an in-vivo environment is vastly different from classical radio frequency (RF) communication environments, and novel communication systems for this environment are needed to be developed. One such system is the molecular communication via diffusion (MCvD) that is based on relaying information over a diffusion channel using special molecules, called messenger molecules (MM) \cite{Farsad2016_Comprehensive}.

In the literature, the performance of an MCvD system is generally evaluated using either the signal-to-noise ratio (SNR), bit-error-rate (BER), or channel capacity. Less prominently, other metrics such as symbol-error-rate, signal-to-interference-noise-ratio (SINR), channel impulse response, and channel capacity considering transmitter energy budget have also been used. Previous studies which consider symbols that are representing multiple bits of information, utilize symbol-error-rate instead of BER \cite{Singhal2015_Performance, Kim2012_Novel}. Other works that take into account the effect of interfering sources (e.g., inter-symbol interference (ISI), co-channel interference, adjacent channel interference) over the system use SINR instead of SNR \cite{Raut2016_Connectivity, Jamali2017_Design}. In \cite{Mahfuz2010_CharOB, Wang2014_Transmit, Chou2012_Molecular}, the authors use channel impulse response to show the effectiveness of the proposed signal shaping method. Lastly, channel capacity can also be expanded to include the energy limitation of the transmitter \cite{Kuran2010_Energy}.

Among the three main metrics mentioned above, BER and channel capacity are defined in the context of MCvD. To the best of our knowledge the physical meaning of SNR and its calculation is not elaborated in detail in this new domain. In RF communication, as an alternative to SNR and SINR, a plot called eye diagram is also used to evaluate the performance of a signal. Different features of this diagram are used to measure various signal characteristics (i.e., eye opening for noise, eye width for jitter, and eye closure for ISI).

In this work, we propose a new diagram called MOL-Eye for evaluating the performance of a molecular signal. Specifically, we propose three performance metrics based on the MOL-Eye diagram as maximum eye height, standard deviation of the number of received molecules, and counting SNR to evaluate the performance of an MCvD system. We evaluate the validity of these performance metrics by comparing the performance of the classical binary concentration shift keying (BCSK) technique \cite{Kim2012_Novel, Kuran2011_Modulation} with an advanced modulation technique we call consecutive power adjustment (BCSK-CPA). BCSK-CPA is based on the power adjustment (BCSK-PA) technique proposed in previous works in the literature \cite{Einolghozati2011_Capacity,Tepekule2014_Energy}. According to our simulation results conducted in a 3D vessel-like environment, the three metrics proposed in the paper successfully depict the advantage of BCSK-CPA over BCSK, which shows the validity of the proposed metrics in the context of MCvD. Even though we consider a vessel-like environment, the eye diagram and the proposed performance metrics can also be applied to free diffusion environments.

The main contributions of the paper are summarized below:
\begin{itemize}
\item Based on the eye diagram concept in classical RF communications, we propose the MOL-Eye diagram in the context of MCvD.
\item Using the MOL-Eye diagram, we introduce three new performance evaluation metrics the performance analysis of a molecular signal, i) maximum eye height-- MaxEH, ii) standard deviation of the number of received molecules, and iii) counting SNR-- CSNR. 
\item We verify the applicability of these three metrics by comparing the performances of BCSK-CPA and basic BCSK techniques in MCvD.
\end{itemize}

\section{System Model}
Most prior work in the molecular communication literature considers a free diffusion environment where the MMs can roam freely without any boundaries (except the transmitter and the receiver) in the communication environment. In contrast, we consider a cylindrical vessel-like environment with a positive flow towards the receiver in this work. This vessel-like environment is more suitable to significant \emph{in-vivo} and \emph{in-vitro} environments, i.e., blood vessels in a human body and micro-fluidic channels.

\subsection{Diffusion Model}

\begin{figure}[!t]
	\begin{center}
	\includegraphics[width=0.98\columnwidth,keepaspectratio]{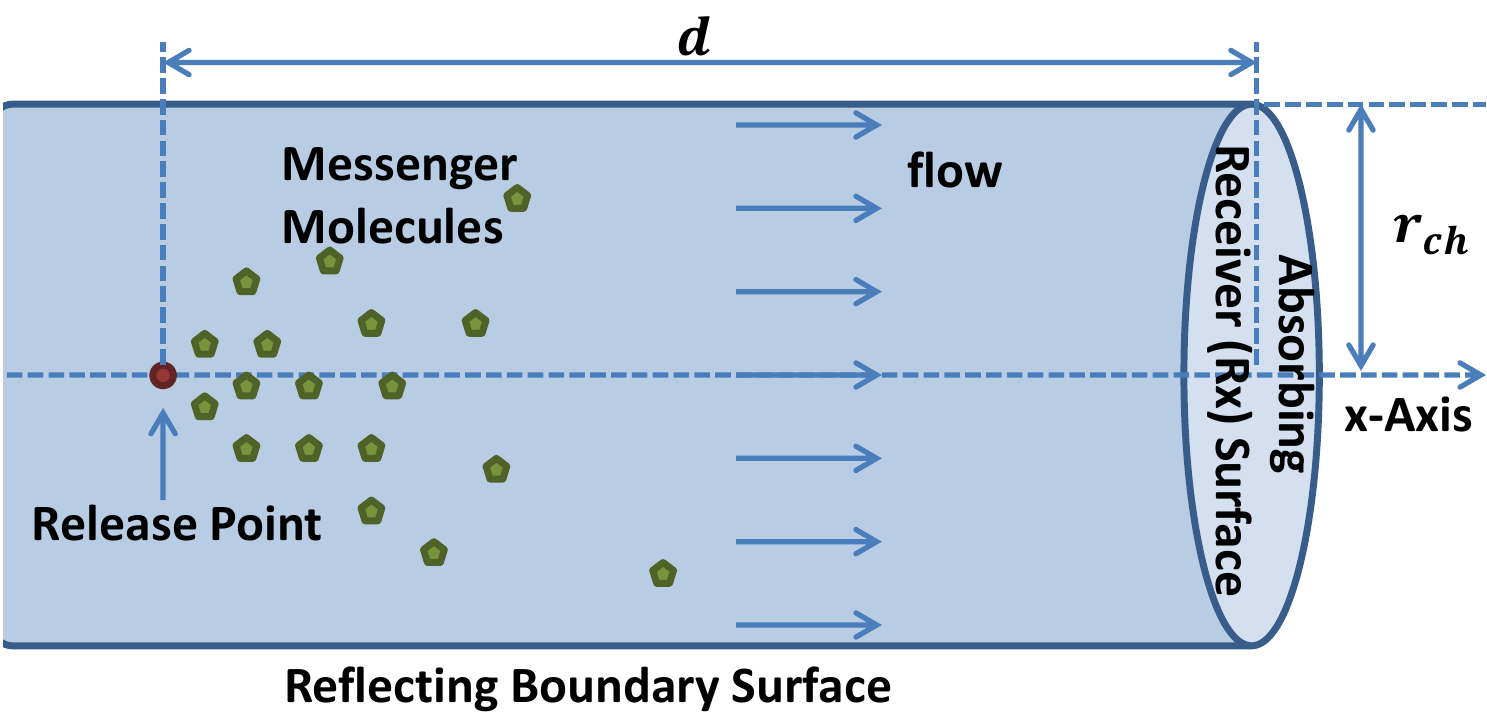}
	\caption{Micro-fluidic based communication channel model representation}
	\label{fig_communication_model}
	\end{center}
\end{figure}

We consider a diffusion model consisting of a point transmitter, a fully absorbing circular receiver, a single type of information carrying MM, and a vessel-like environment with laminar flow. The vessel-like environment is considered to be a perfect cylinder with a complete reflecting surface (Fig.~\ref{fig_communication_model}). Since this is a closed environment with a positive flow towards the receiver, the surviving probability of MMs is much lower than the unbounded environment case (i.e., more MMs hit the receiver).

In the diffusion model, the total displacement along the x-axis ($\deltaX$) of an MM in $\deltat$ duration is calculated as the sum of the displacement due to the flow ($\deltaXsub{flow}$) and displacement due to the diffusion ($\deltaXsub{diffusion}$) as 
\begin{align}
\begin{split}
\deltaX &= \deltaXsub{flow} + \deltaXsub{diffusion} \\
        &= \vflow \, \deltat + \deltaXsub{diffusion}
\end{split}
\label{eq_displacementX}
\end{align}
where $\vflow$ is the laminar flow velocity.

Similar to the classical diffusion model without flow, the displacement due to diffusion follows a Gaussian distribution 
\begin{align}
\deltaXsub{diffusion} \sim \mathscr{N}(0, 2D\Delta t)
\label{eq_displacementX_diffusion}
\end{align}
where $\Delta t$ is the simulation time step, $\diCO$ is the diffusion coefficient,  and $\yakinsama$ is the Gaussian random variable with mean $\mu$ and variance $\sigma^2$. Considering the movement in all three axes, the total displacement in a single time step is calculated as:
\begin{align}
\rarrow = (\deltaX, \deltaY, \deltaZ)
\label{eq_displacementXYZ}
\end{align}
where $\deltaY$ and $\deltaZ$ correspond to the displacement in the y- and the z-axes, respectively, both of which follow a Gaussian distribution with the same $\mu$ and $\sigma$ values with $\deltaXsub{diffusion}$.

\subsection{Modulation and Demodulation}
In this study, we use binary concentration shift keying (BCSK) as the modulation technique with a symbol duration of $\symdur$~\cite{Kuran2011_Modulation}. In BCSK, a given symbol at the \(k^{th}\) symbol duration (i.e., \(\symbolTx{k}\))  can either represent bit-0 or \bitone{} (i.e., \(\symbolTx{k} \in \{0, 1\} \)). Based on this value, the transmitter releases \(\ntxk{k}\) MMs where \(\ntxk{k}=n_{S[k]}\). In order to increase the detectability of bit-0s and \bitones{}, we choose \(\ensayisi{0}\) as 0 and \(\ensayisi{1}\) as the minimum number of molecules sufficient to have a smooth communication.

At the receiver side, \(\nrxk{k}{}\) represents the number of MMs arriving at the receiver within the \(\nth{k}\) symbol slot, which includes both MMs from the current symbol and the previous symbols. As in (\ref{eq:CSK_Rx_Decode}), the receiver applies a basic thresholding on \(\nrxk{k}{}\) to decode the signal (\(\symbolRx{k}\)) as either bit-0 or \bitone{}.
\begin{equation} \label{eq:CSK_Rx_Decode}
\symbolRx{k}= 
  \begin{cases}
    0, & \nrxk{k}{} <  \lambda \\
    1, & \nrxk{k}{} \geq \lambda \\
  \end{cases}
\end{equation}

In addition to the basic BCSK technique we have also implemented a variant of BCSK with power adjustment (BCSK-PA) technique proposed in \cite{Tepekule2014_Energy} that we call BCSK with consecutive power adjustment (BCSK-CPA). BCSK-PA focuses on minimizing the variation between the \(\nrxk{k}{}\) values where \(\symbolTx{k}\!=\! 1\), regardless of the values of the past $m$ symbols by regulating the molecular emission rate. By doing so, BCSK-PA aims to considerably reduce the effect of ISI in MCvD. To this end, in BCSK-PA the transmitter uses emission rates based on past symbol values as \(\shistorykm{k}{m}\) where \(\shistorykm{k}{m}=(\symbolTx{k-1}, \symbolTx{k-2},...,\symbolTx{k-m})\) is a vector representing the symbol values of the previous \(m\) symbols at the $\nth{k}$ symbol slot (i.e., the history of bits at the $\nth{k}$ symbol slot).

\begin{figure}[t]
	\begin{center}	\includegraphics[width=0.98\columnwidth,keepaspectratio]{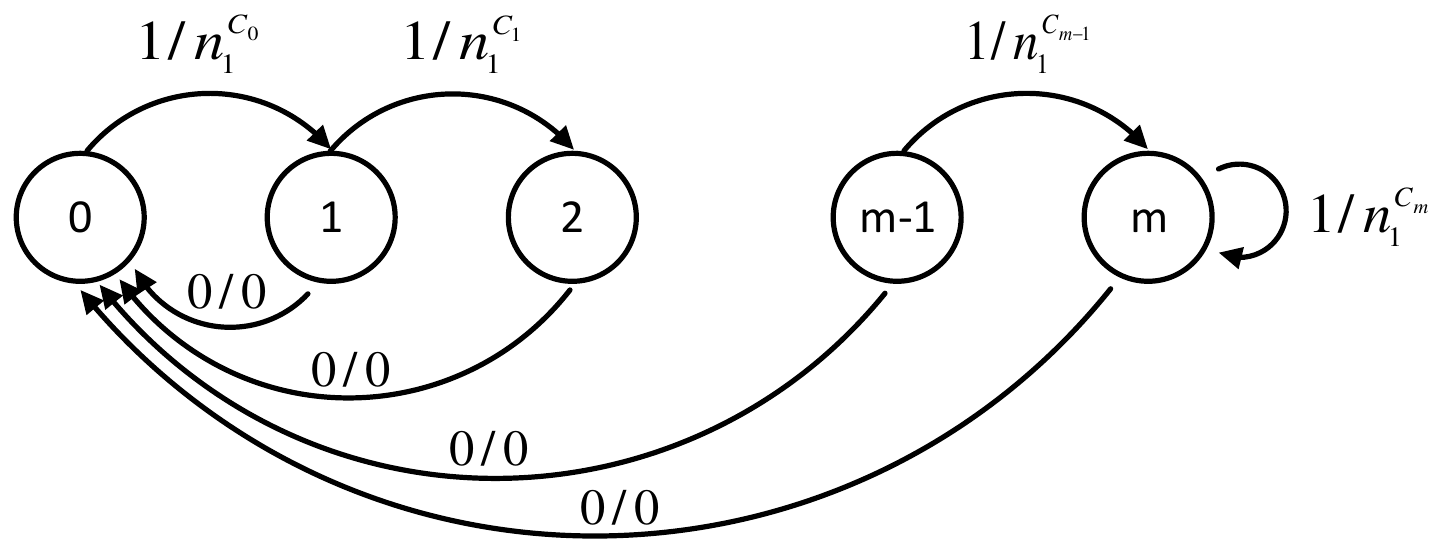}
	\caption{State diagram of BCSK-CPA. The state diagram counts the consecutive \bitones{} until the current bit. State transitions are given in the form of $r/o$, where $r$ and $o$ represent \(\symbolTx{k}\) and the \(\ntxk{k}\), respectively (e.g., if \(\symbolTx{k} = 1\) when the state is 1, the new state becomes 2 and $\cpantx{1}{1}$ molecules will be emitted in the current symbol slot).}
	\label{fig:csk_cpa}
	\end{center}
\end{figure}

Although BCSK-PA reduces the ISI effect in the communication, it requires a considerable amount of memory at the transmitter side, especially as \(m\)  increases (e.g., BCSK-PA with \(m=10\) requires 1024 distinct \(\shistorykm{k}{m}\) cases and corresponding emission amounts). Our new technique, BCSK-CPA, aims reducing this memory requirement by only considering the cases where the previous symbols have consecutive \bitones{} to change the emitted MM count. Fig.~\ref{fig:csk_cpa} shows the state diagram of a transmitter utilizing BCSK-CPA with $m$-memory where the state number represents the consecutive \bitones{}. Also, \(\cpahistory{k}\) replaces \(\shistorykm{k}{m}\) from CSK-PA and denotes the number of consecutive \bitones{} just before the $\nth{k}$ symbol slot as
\begin{equation} \label{eq:CSK_CPA_States}
\cpahistory{k}= 
  \begin{cases}
    0, & \symbolTx{k\!-\!1} \!= 0 \\
    1, & \symbolTx{k\!-\!2} \!= 0 \, ,\, \symbolTx{k\!-\!1} \!= 1 \\
    2, & \symbolTx{k\!-\!3} \!= 0 \, ,\, \symbolTx{k\!-\!2} \!= 1 \,,\, \symbolTx{k\!-\!1} \!= 1 \\
    . & . \\
    m, & \forall i \; k\!-\!m\leq i < k \;\symbolTx{i} \!= 1
  \end{cases}
\end{equation}

The rationale behind BCSK-CPA is the fact that in a BCSK system, the effect of bit-0s over ISI is much less than \bitones{}. Therefore, we can omit the effect of bit-0s in the past symbol values to reduce the memory requirements of the technique while not considerably impairing the performance of the system.

When molecules are emitted from the emission point, some of them hit the receiver in the current symbol slot while the rest resides in the channel and can be received during the successive symbol slots. We define $\pk{i}$ as the mean fraction of emitted molecules that are received during the $\nth{i}$ following symbol slot. Please note that $\pk{0}$ corresponds to the mean fraction of molecules to be absorbed during the current symbol slot. Therefore, the expected number of molecules to be absorbed in the $\nth{k}$ symbol slot becomes 
\begin{align}
\mathbf{E}(\nrxk{k}{}) = \pk{0} \ntxk{k}  + \underbrace{\sum_{i=1}^{m} \pk{i} \ntxk{k - i}}_{\text{residual}}
\label{eqn_expected_arrival}
\end{align}
where $\mathbf{E}(\cdot)$ is the expectation operator and $\isiwin$ is the number of past symbols that are assumed to be affecting the current symbol, which is also called as the ISI window length.

For BCSK-CPA, the number of molecules to emit is adjusted according to the number of consecutive \bitones{} as explained above. Therefore, the expected number of residual molecules ($\nresidual{k}$) for the $\nth{k}$ symbol slot becomes
\begin{align}
\mathbf{E}(\nresidual{k}) = \sum_{i=1}^{\cpahistory{k}} \pk{i} \ntxk{k - i}.
\label{eq_residual} 
\end{align}
Please note that $\mathbf{E}(\nresidual{k})$ can be calculated by the transmitter since the transmitted bits are known perfectly. Hence, the number of molecules to emit is adjusted as
\begin{align}
\ntxk{k} = 
  \begin{cases}
    \ensayisi{0}                      & \symbolTx{k} \!= 0 \\
    \cpantx{k}{1} = \cpantx{0}{1} - \frac{\mathbf{E}(\nresidual{k})}{\pk{0}}  & \symbolTx{k} \!= 1
  \end{cases}
\label{eq_transmitted}
\end{align}
where $\cpantx{k}{1}$ denotes the number of molecules to emit when the number of consecutive \bitones{} is $\cpahistory{k}$. Note that this scheme ensures to have approximately the same expected number of molecules (at the receiver side) for each \bitone{}, namely $\pk{0} \, \cpantx{0}{1}$, which is equal to $\pk{0} \, \ensayisi{1}$.

\subsection{BER Formulation with and without CPA}
In \eqref{eqn_expected_arrival}, the expected number of received molecules at symbol slot $k$ is given, and $\nrxk{k}{}$ exhibits a binomial random variable~\cite{yilmaz2015arrivalMA}. For tractability, we approximate the binomial random variables with a Gaussian random variable as 
\begin{align}
\begin{split}
\label{eqn_arrival_rv}
\nrxk{k}{} &\sim \mathscr{N}(\mu_k, \sigma^2_k) \\
\mu_k        &=\sum_{i=0}^{m} \pk{i}\, \ntxk{k-i} \\
\sigma^2_k   &=\sum_{i=0}^{m} \pk{i} \,(1-\pk{i})\, \ntxk{k-i} \\
\end{split}
\end{align}
where $\ntxk{k-i}$ differs for BCSK-CPA and BCSK. We acquire the $\pk{i}$ values by simulation. Then, we can evaluate the probability of error by using Gaussian distribution tail probabilities. 

Considering the history that includes the previous bits, we obtain the error probability at the $\nth{k}$ symbol slot as
\begin{align}
\begin{split}
\proberr   &= \proberrsub{\symbolTx{k}\!=0, \shistorykm{k}{k\!-\!1}} \, \prob(\symbolTx{k}\!=0, \shistorykm{k}{k\!-\!1}) \\
		   &+ \proberrsub{\symbolTx{k}\!=1, \shistorykm{k}{k\!-\!1}} \, \prob(\symbolTx{k}\!=1, \shistorykm{k}{k\!-\!1})
\end{split}
\end{align}
where $\proberrsub{\symbolTx{k}, \shistorykm{k}{k\!-\!1}}$ corresponds to the probability of error given that the history bits are $\shistorykm{k}{k\!-\!1}$ and the current bit is $\symbolTx{k}$. We evaluate these probabilities with the tail probabilities of the approximate arrival distribution (i.e., the random variable in \eqref{eqn_arrival_rv}) and considering only the ISI window (i.e., $\shistorykm{k}{m}$). Please note that we do not need all of the previous bit values to implement BCSK-CPA. However, we need these bit values for the evaluation of BER for analysis purposes.
\begin{figure}[!t]
\begin{center}
\includegraphics[width=0.98\columnwidth,keepaspectratio]{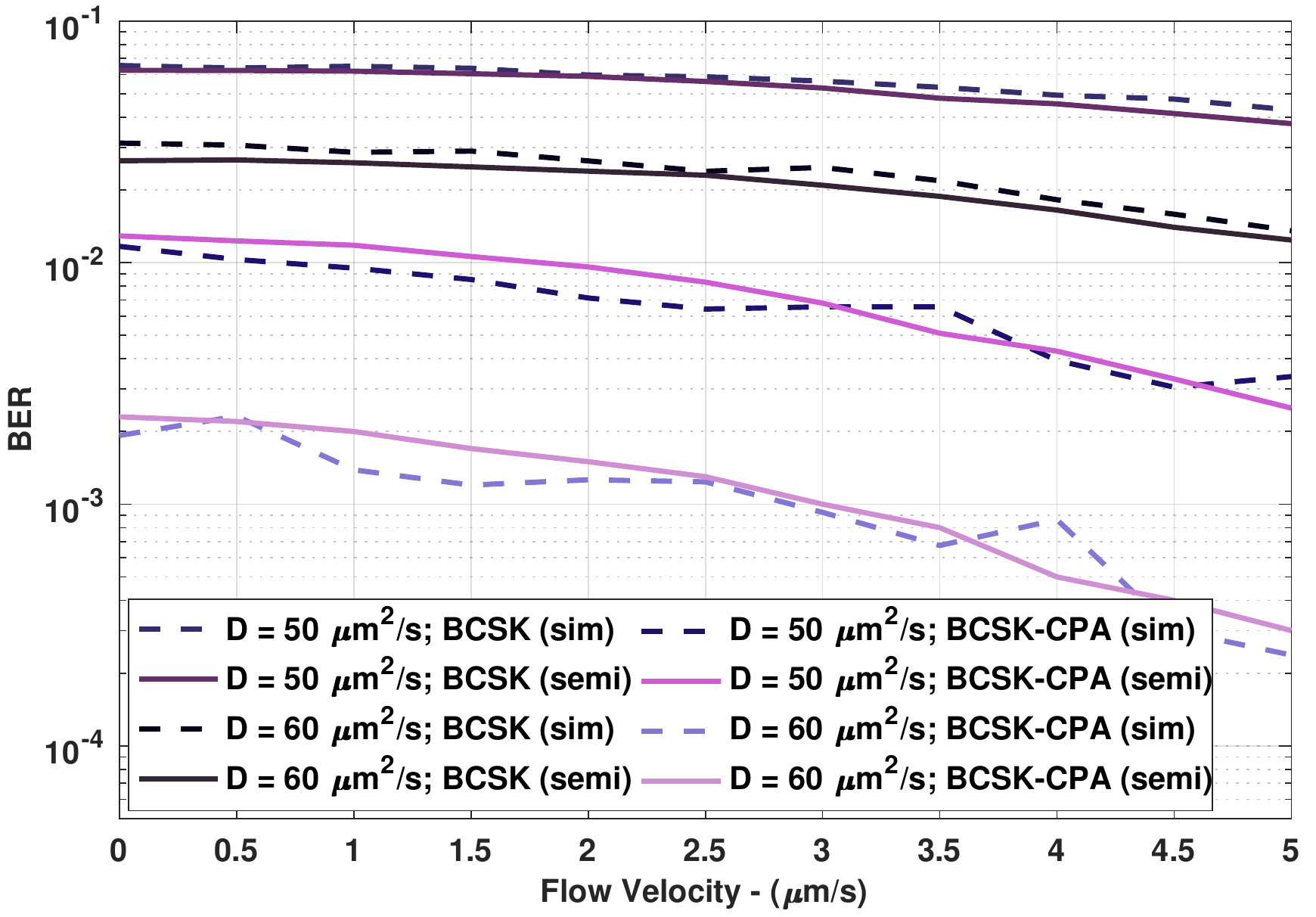}
\caption{BER plot for BCSK and BCSK-CPA. Curves with the name \textit{sim} and \textit{semi} correspond to the simulation and the semi analytical method. ($d=\SI{6}{\micro\metre}$, $\ensayisi{1}=300$, $\symdur=\SI{0.4}{\second}$)} 
\label{fig_validation_flow_vs_BER}
\end{center}
\end{figure}

In Fig.~\ref{fig_validation_flow_vs_BER}, we plot the flow velocity versus BER values for the two modulation techniques (i.e., BCSK and BCSK-CPA) considering two different $\diCO$ values. In all cases, the simulation and analytical method values are coherent with each other (i.e., simulation results are validated by the analytical method values). Moreover, as expected BCSK-CPA outperforms BCSK by a considerable margin. Additionally, the results show that BER and laminar flow is inversely proportional to one another.

\section{Eye Diagram and MOL-Eye Diagram}
Eye diagram, also called as eye pattern, is a method for measuring the quality of a signal \cite{Freude2012_Quality}. The name of eye diagram comes from its shape. The width of the eye defines the time interval of the received signal without ISI. Therefore, the more the eye is open, the less the ISI level is, and vice versa.


Eye diagrams, obtained by using oscilloscope, are mostly used by field engineers. Eye diagram is useful for detecting problems such as noise, jitter, and attenuation. The conventional eye diagram has five metrics that are also applicable in MC:
\begin{itemize}
\item \textit{0 and 1 level}: The mean values of bit-0 and \bitone{} curves in the diagram (dashed lines in Fig.~\ref{fig_eye_diagrams}).
\item \textit{Rise and fall time}: Transition times of the data to the upward and downward slope of the eye diagram.
\item \textit{Eye amplitude}: The biggest distance between the mean of bit-0 and the mean \bitone{} curves.
\end{itemize}

In the context of molecular communication (MC), we propose MOL-Eye as  analogous to the eye diagram in conventional communications. MOL-Eye is a good way to visualize signals in MC, as can be seen in Fig.~\ref{fig_eye_diagrams}. To obtain the eye diagrams in the figure, the received signals of consecutive bit transmissions are repetitively sampled and applied in an overlapping fashion. Moreover, curves of \bitone{} transmissions are in dark blue whereas bit-0s' are in light blue. The diagrams in the first row are generated using the BCSK technique, whereas the diagrams in the second row are generated using the BCSK-CPA technique. Also, environmental conditions get worse from left to right, and consequently the openness of the MOL-Eye starts to decrease.

In the MC literature, BER is used extensively to measure the quality of the signal. However, BER calculation requires excessive processing power, which would be unsuitable for nanomachines that are expected to have very low energy budgets. Therefore, we propose three performance metrics derived from MOL-Eye diagram as alternative performance metrics. We use the conventional eye diagram metric called, maximum eye height (MaxEH), as well as propose two new metrics for the molecular signal, namely the standard deviation of the received molecules and counting SNR (CSNR). Especially, CSNR is a promising metric since we observe a one-to-one relation between CSNR and BER. Therefore, if the relation between BER and CSNR can be formulated, BER evaluation and optimization process will be much more efficient.

In this work, we propose CSNR as a supportive metric to BER. To calculate CSNR, we first define the integral difference between every combination of \bitone{} and bit-0 curves as in \eqref{eq_integralDif}
\begin{align}
\intdiff(i,j) =  \int_{0}^{\symdur} c_1(i) -  c_0(j) \,dt
\label{eq_integralDif}
\end{align}
where $c_1(i)$ and $c_0(j)$ are the $\nth{i}$ \bitone{} and $\nth{j}$ bit-0 sampled curves, and $\symdur$ is the symbol duration. Consequently, we calculate the mean, $\mu_{\intdiff}$, and the standard deviation, $STD(\intdiff)$, of $\intdiff$ values. Finally, we calculate CSNR as in
\begin{align}
CSNR = \frac{\mu_{\intdiff}}{STD(\intdiff)},
\label{eq_CSNR}
\end{align}
which is an alternative definition of SNR for non-negative signals \cite{Schroeder2000_Astronomical}.

\section{Numeric Results}
The results presented in this section are obtained from the custom-made simulator that keeps track of the hitting molecules, which is the number of successfully received molecules in every simulation time step. By utilizing the simulation output, we evaluate the aforementioned eye diagram metrics and BER values under different conditions. 

We consider MCvD in a vessel-like environment as depicted in Fig.~\ref{fig_communication_model}. The system parameters are given in Table~\ref{simulation_parameters}.

\begin{table}[th]
\renewcommand{\arraystretch}{1.2}
\caption{Simulation parameters}
\label{simulation_parameters}
\begin{center}
\begin{tabular}{ L{5cm}  L{3cm} }
  \hline			
  Parameter & Value \\
  \hline			  
  Radius of channel ($r_{\text{ch}}$)& \SI{5}{\micro\metre} \\
  Radius of the receiver & \SI{5}{\micro\metre} \\
  Distance between Tx and Rx ($d$) & $\{4, 5, 6\} \si{\micro\meter}$ \\
  Diffusion coefficients ($\diCO$) & $\{50, 100, 150 \}\si{\micro\metre^{2}\per\second}$ \\
  Laminar flow velocities ($\vflow$)& $0\sim 5 \si{\micro\metre/\second}$ \\
  Simulation time step ($\deltat$)    & \SI{0.1}{\micro\second} \\
  $\ensayisi{1}$ (BCSK), $\cpantx{0}{1}$ (BCSK-CPA)$^a$  & $50 \sim 300$ \\
  Symbol duration ($\symdur$)             & \{0.4, 0.5\}\si{\second} \\
  Length of bit sequence       & 100 bits \\
  Number of replications       & 250 \\  
  \hline  
\end{tabular}
\end{center}
\hfill\footnotesize{$^a$ In the rest of this section, for the sake of simplicity, we refer to both of these parameters simply as $\ensayisi{1}$.}
\end{table}

\subsection{BER Analysis}
\begin{figure}[!t]
\begin{center}
\includegraphics[width=0.98\columnwidth,keepaspectratio]{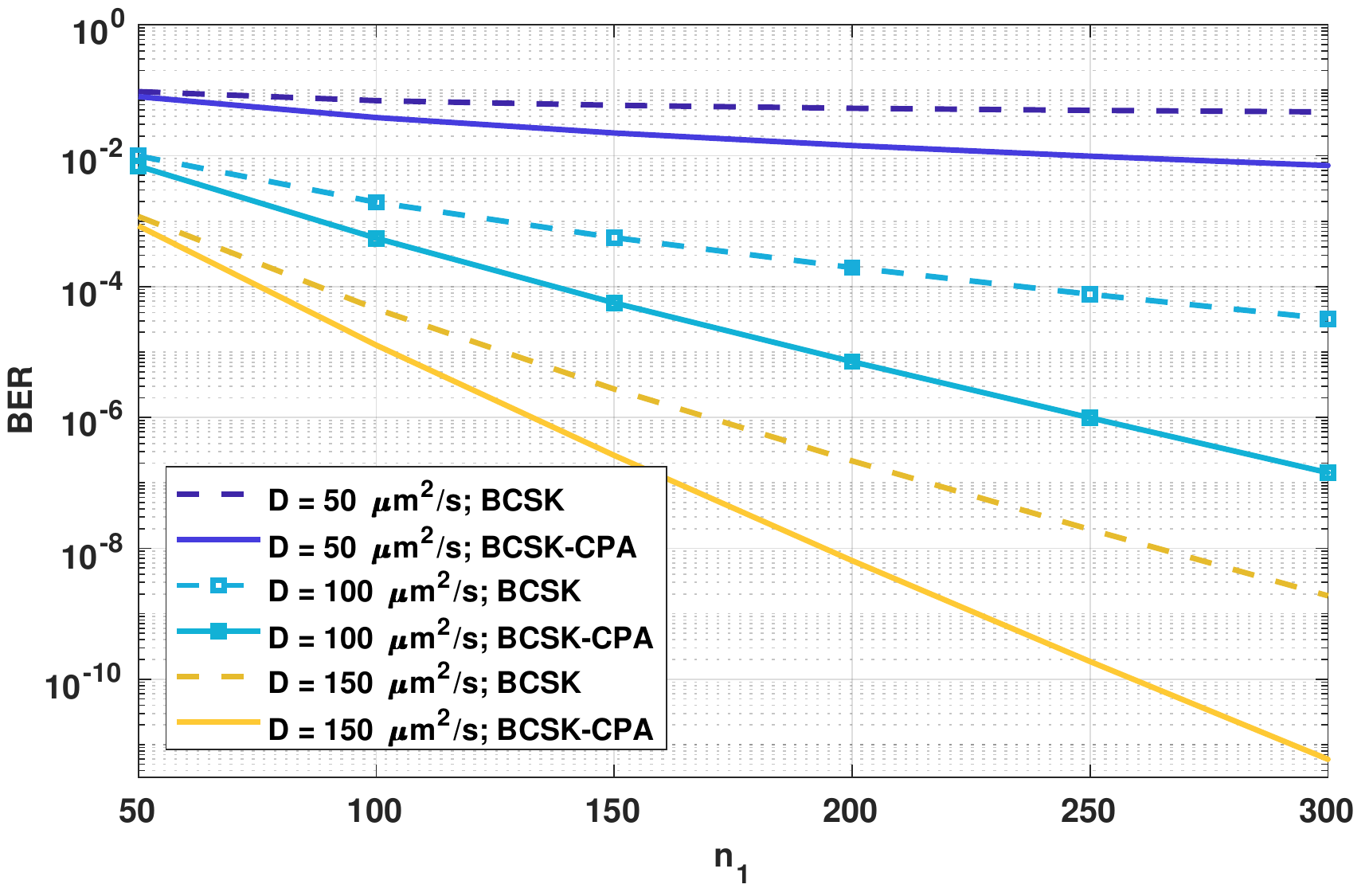}
\caption{Number of molecules vs. BER ($d=\SI{6}{\micro\metre}$, $\vflow=0$, $\symdur=\SI{0.4}{\second}$) }
\label{fig_numOfMol_vs_BER}
\end{center}
\end{figure}

In Fig.~\ref{fig_numOfMol_vs_BER}, BER values corresponding to various $\diCO$ values, $\ensayisi{1}$ values, and modulation techniques are presented. According to the figure, BER decreases as $\diCO$ and $\ensayisi{1}$ increase. Note that the relative gain of BCSK-CPA compared to BCSK is greater for higher $\diCO$ values.

\begin{figure}[!t]
\begin{center}
\includegraphics[width=0.98\columnwidth,keepaspectratio]{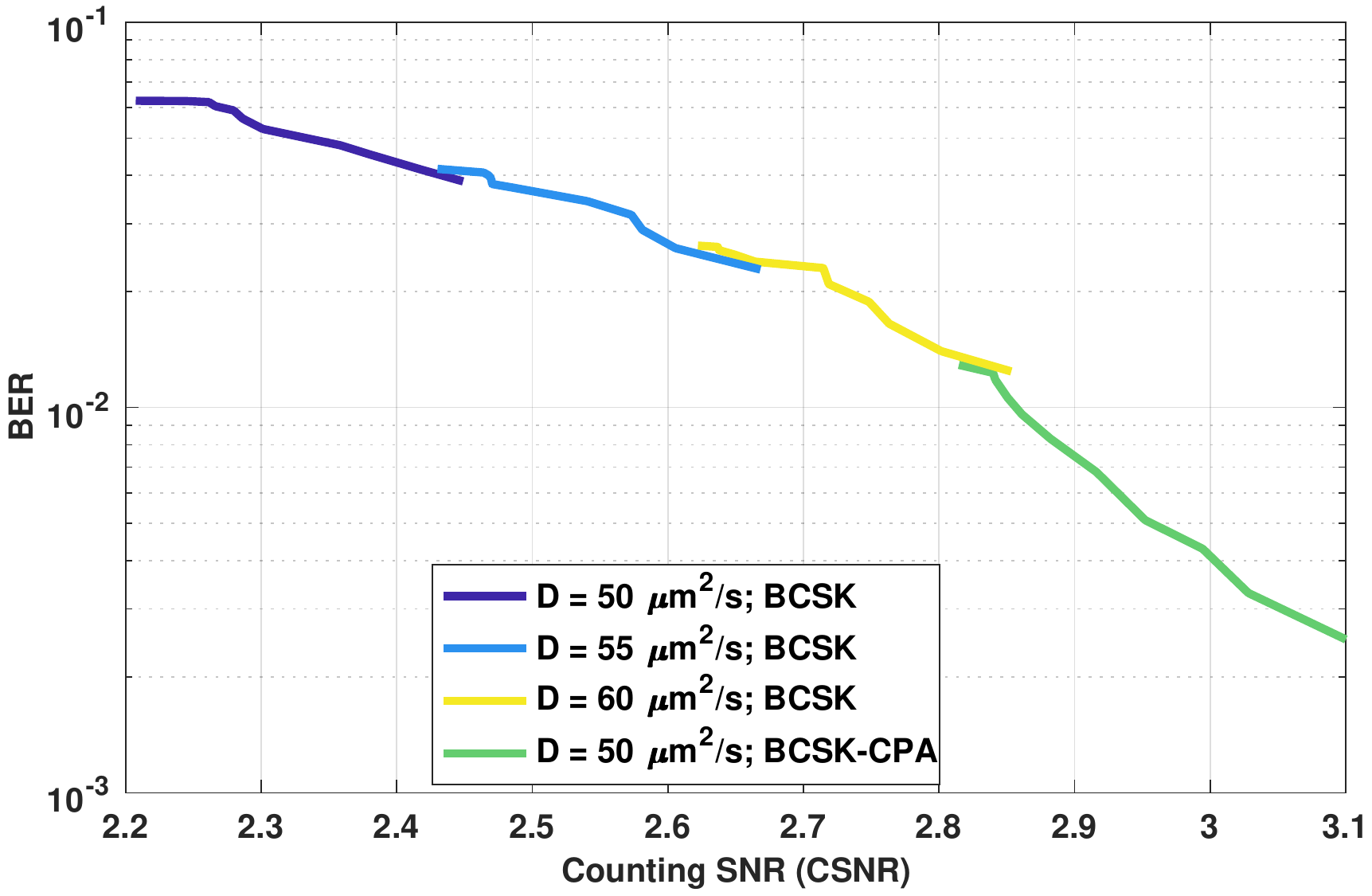}
\caption{CSNR vs. BER ($d=\SI{6}{\micro\metre}$, $\vflow=0$, $\ensayisi{1}=300$, $\symdur=\SI{0.4}{\second}$) } 
\label{fig_CSNR_vs_BER}
\end{center}
\end{figure}

Since CSNR represents the quality of the signal across noise, it is also expected to be inversely proportional to BER. We validate this by running simulations for 11 different flow values from \SI{0}{\micro\metre/\second} to \SI{5}{\micro\metre/\second} that are sequentially increased by \SI{0,5}{\micro\metre/\second}, three different $\diCO$ values, and two different modulation techniques: BCSK and BCSK-CPA (Fig.~\ref{fig_CSNR_vs_BER}). Finally, the relation between CSNR and BER is injective for the given parameters, which means BER can be formulated in terms of CSNR - if the derivations are tractable and can be formulated, BER calculations for CSK-based modulations will be eased. For the future work, we will focus on the analytical derivation of the relation between CSNR and BER.

\begin{figure}[!t]
	\begin{center}
	\includegraphics[width=0.98\columnwidth,keepaspectratio]{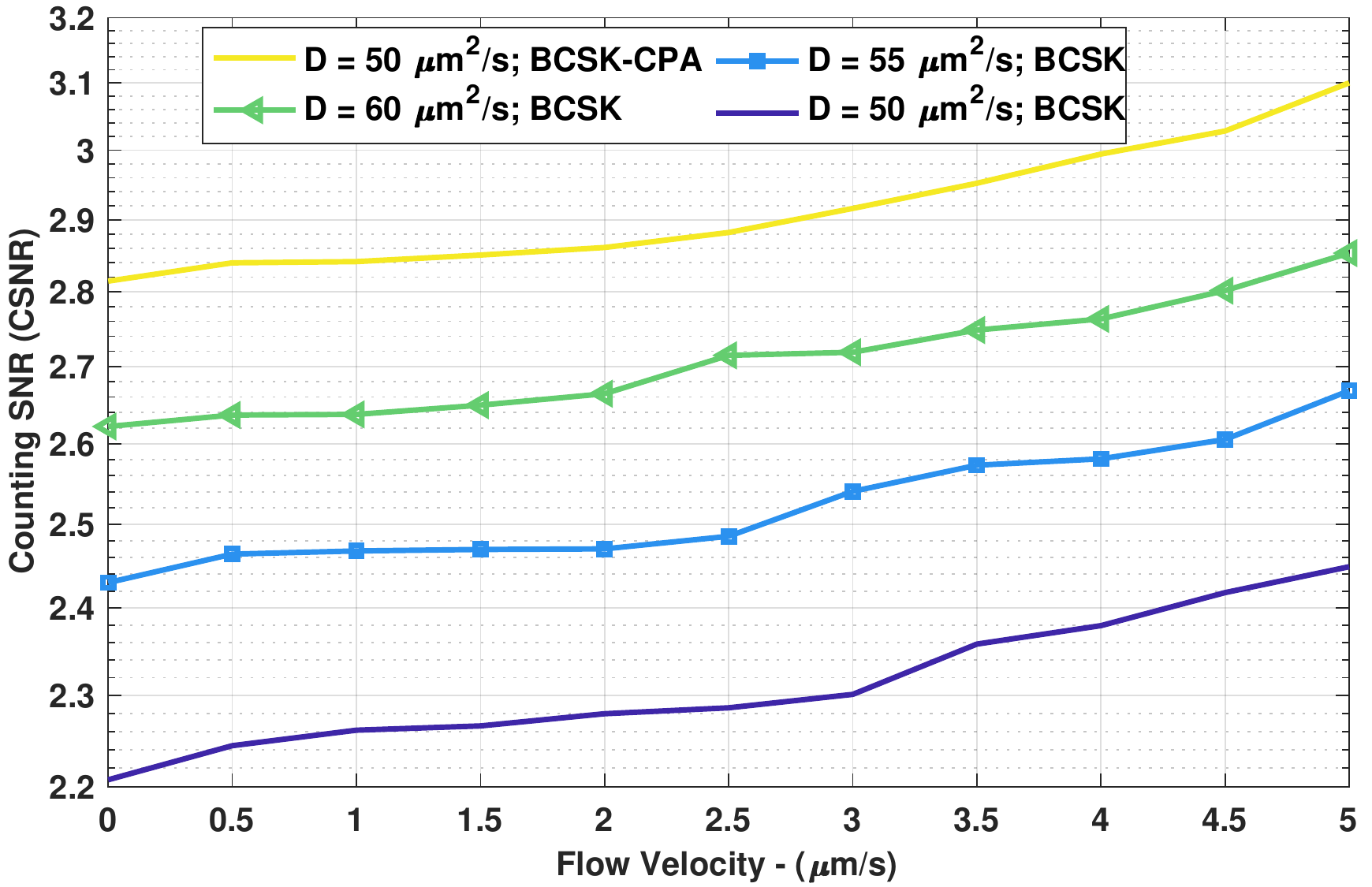}
	\caption{Flow velocity vs. CSNR ($d=\SI{6}{\micro\metre}$, $\ensayisi{1}=300$, $\symdur=\SI{0.4}{\second}$) } 
	\label{fig_flow_vs_CSNR}
	\end{center}
\end{figure}

\begin{figure*}[!t]
	\begin{center}
    \subfigure[Good environment - BCSK]
		{\includegraphics[width=0.66\columnwidth,keepaspectratio]%
		{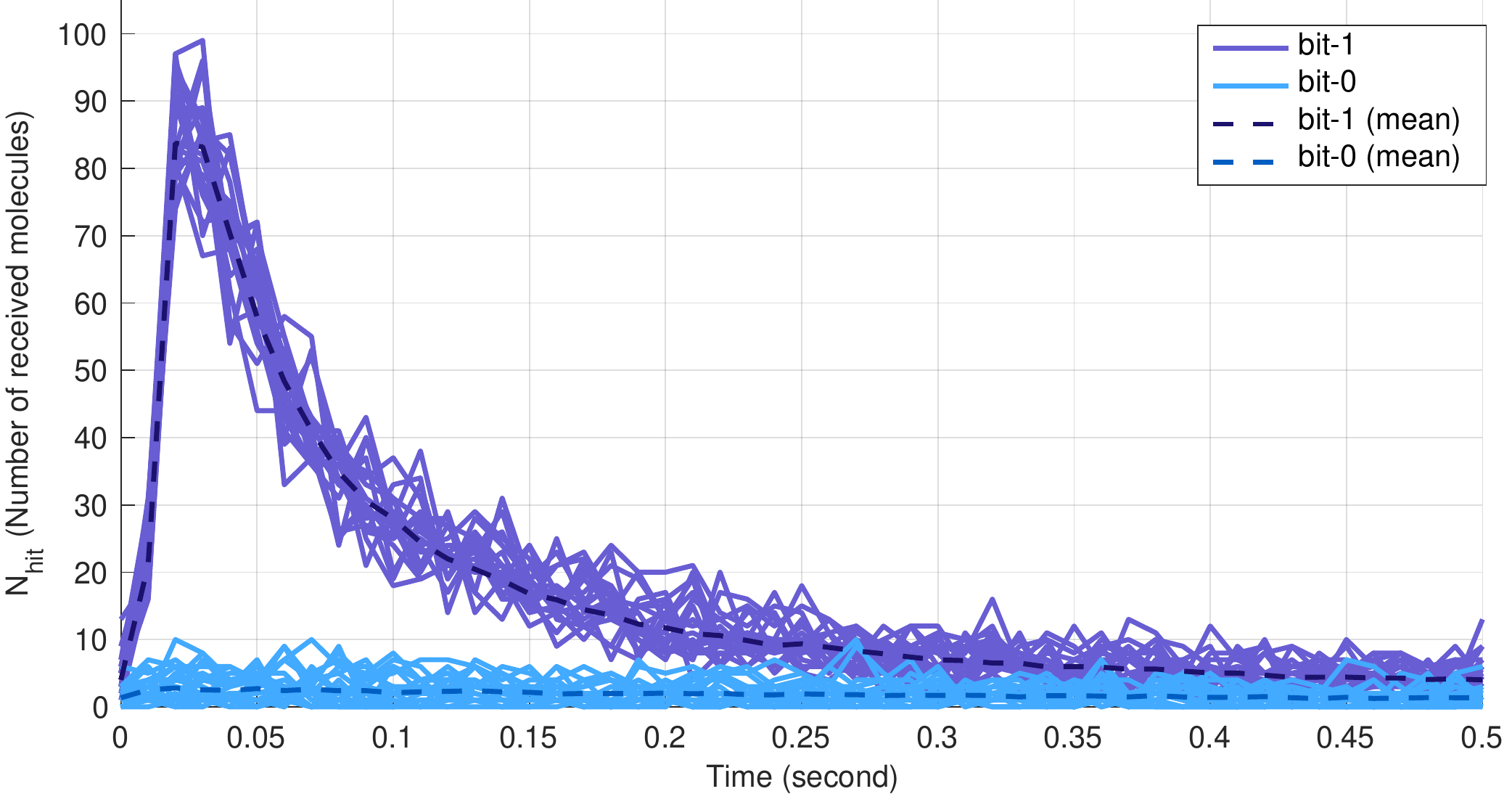}
		\label{fig_eye_diagram_good_woPA}}
    \subfigure[Moderate environment - BCSK]
        {\includegraphics[width=0.66\columnwidth,keepaspectratio]%
		{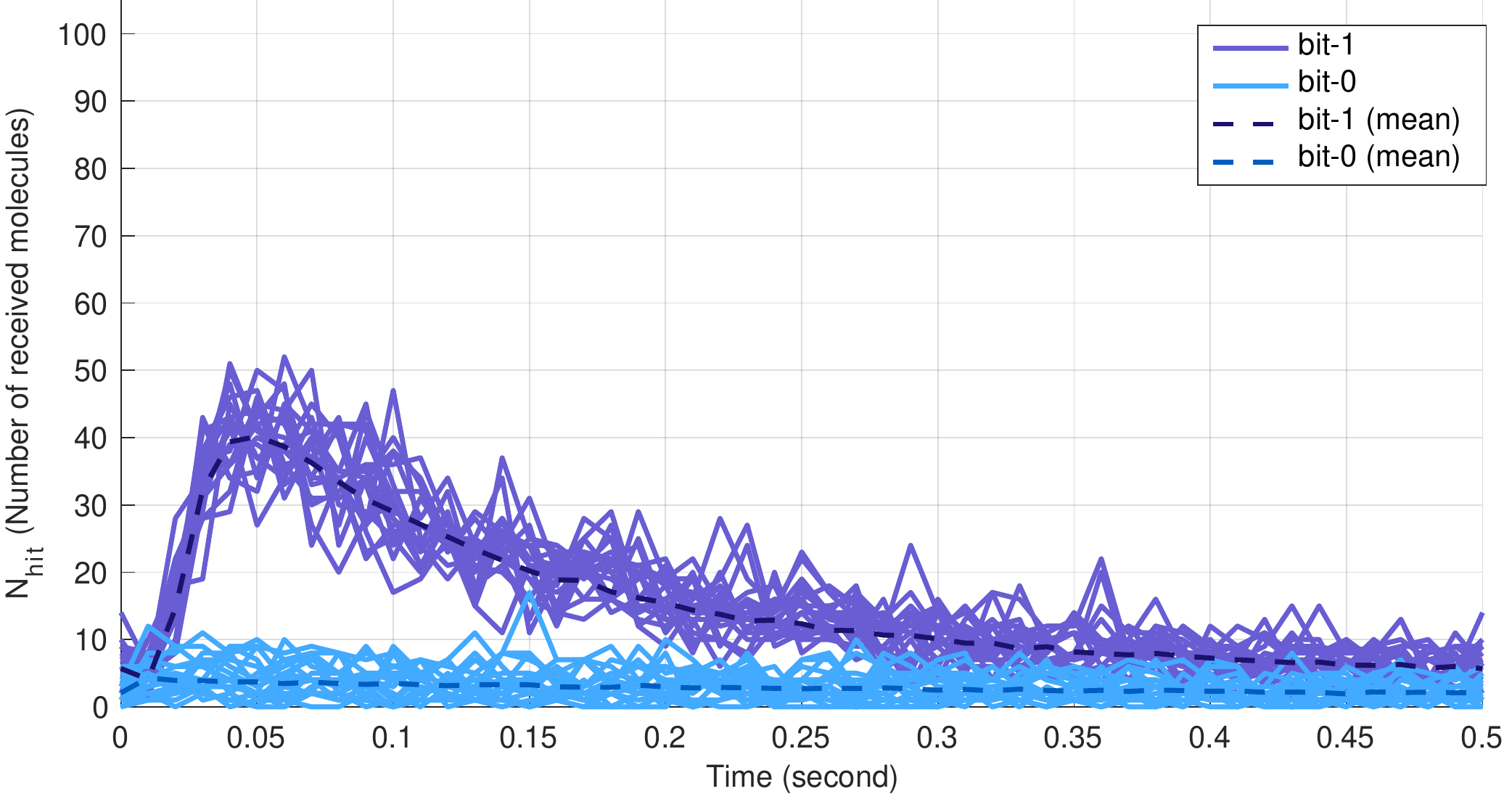}
        \label{fig_eye_diagram_average_woPA}}
    \subfigure[Harsh environment - BCSK]
        {\includegraphics[width=0.66\columnwidth,keepaspectratio]%
		{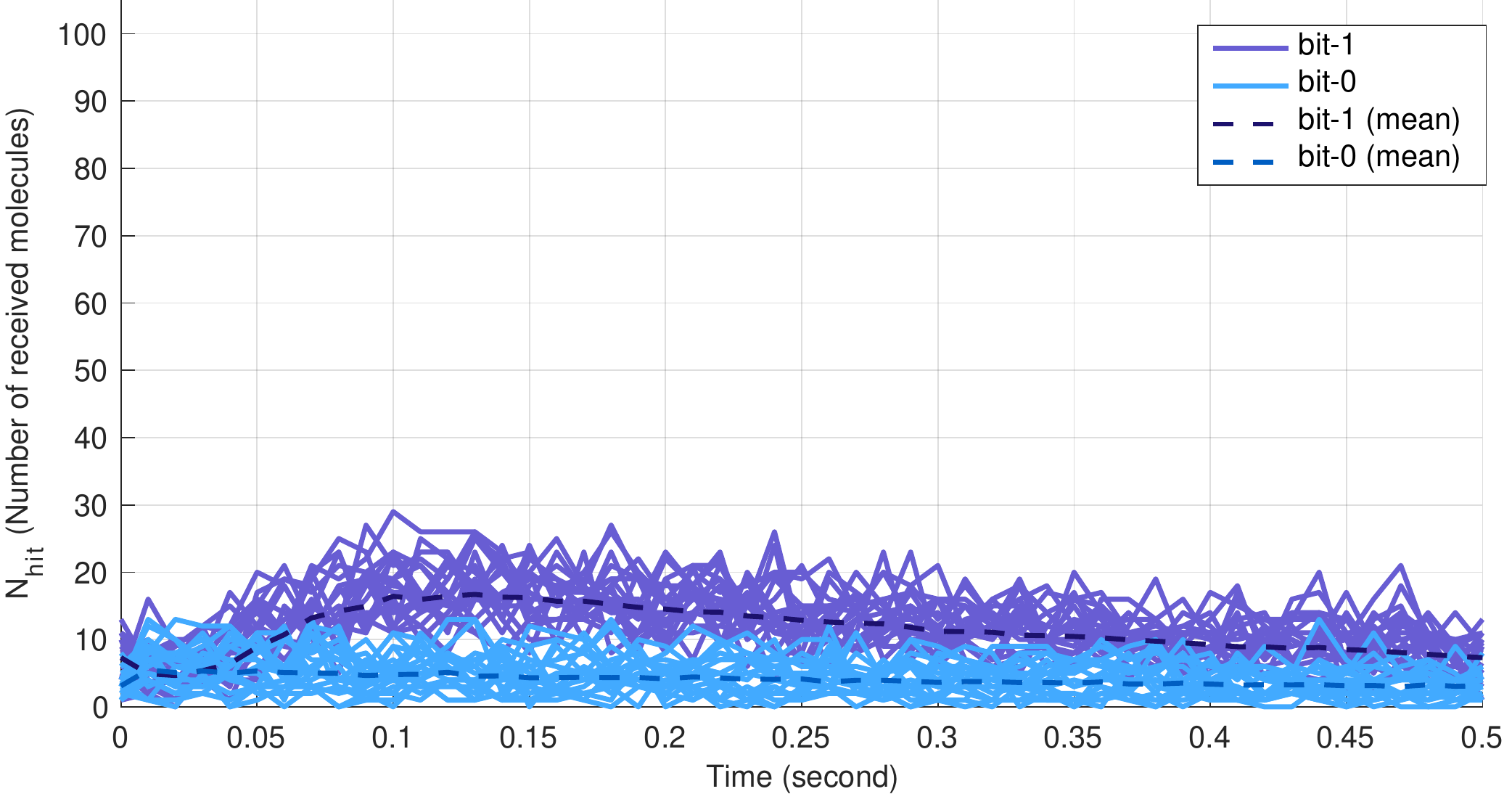}
        \label{fig_eye_diagram_bad_woPA}}        
    \subfigure[Good environment - BCSK-CPA]
		{\includegraphics[width=0.66\columnwidth,keepaspectratio]%
		{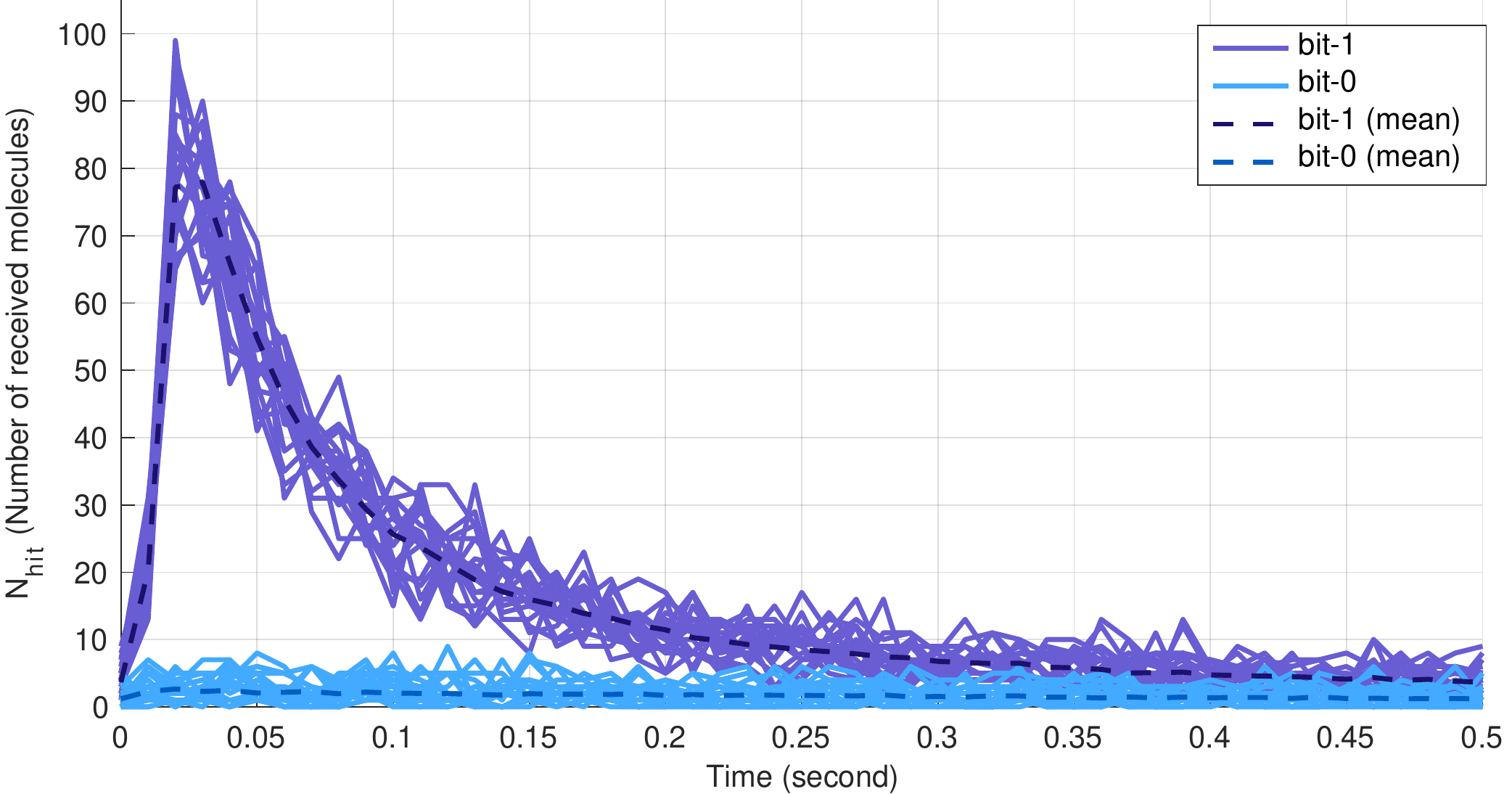}
		\label{fig_eye_diagram_good_withPA}}
    \subfigure[Moderate environment - BCSK-CPA]
        {\includegraphics[width=0.66\columnwidth,keepaspectratio]%
		{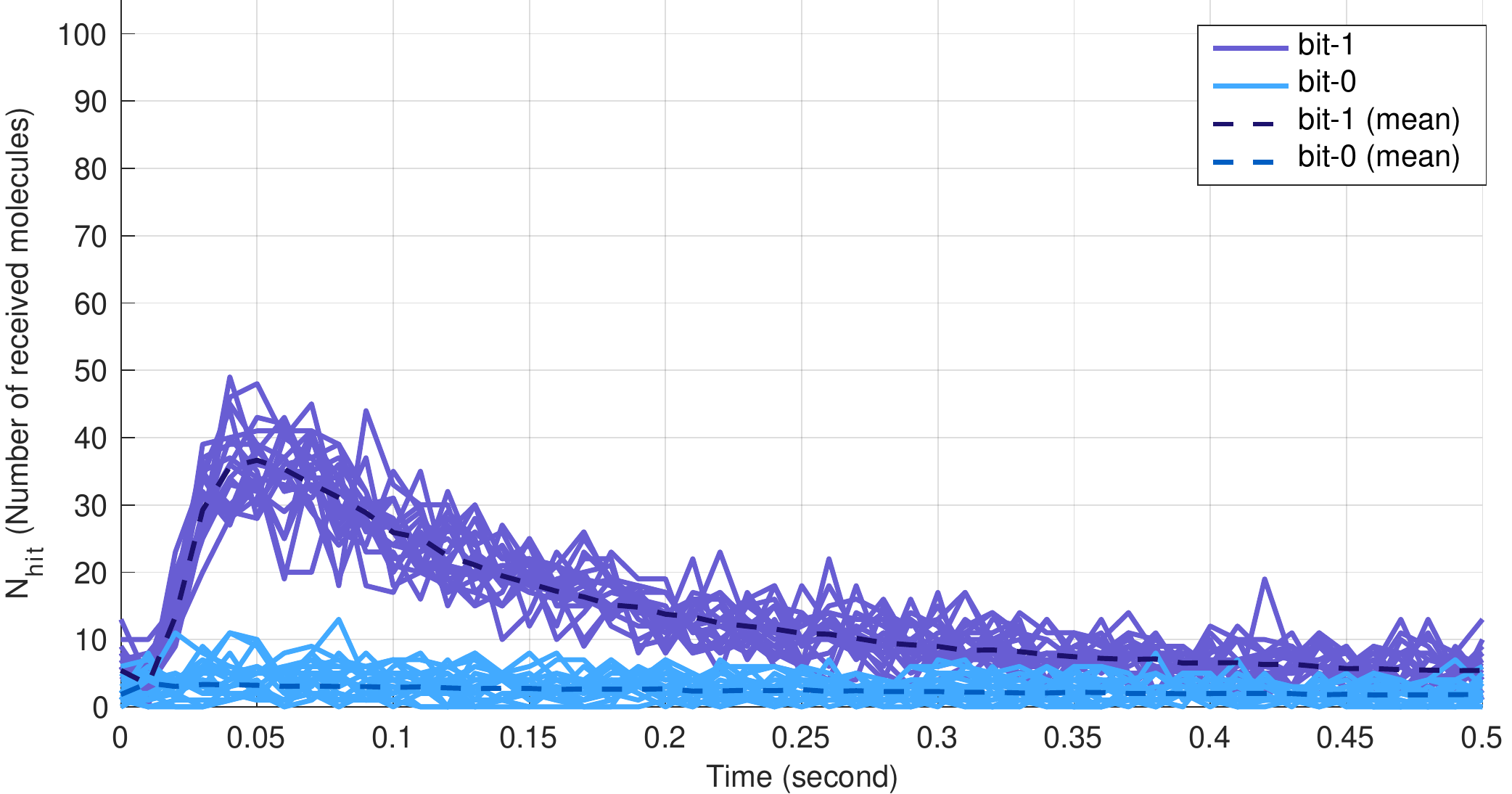}
        \label{fig_eye_diagram_average_withPA}}
    \subfigure[Harsh environment - BCSK-CPA]
        {\includegraphics[width=0.66\columnwidth,keepaspectratio]%
		{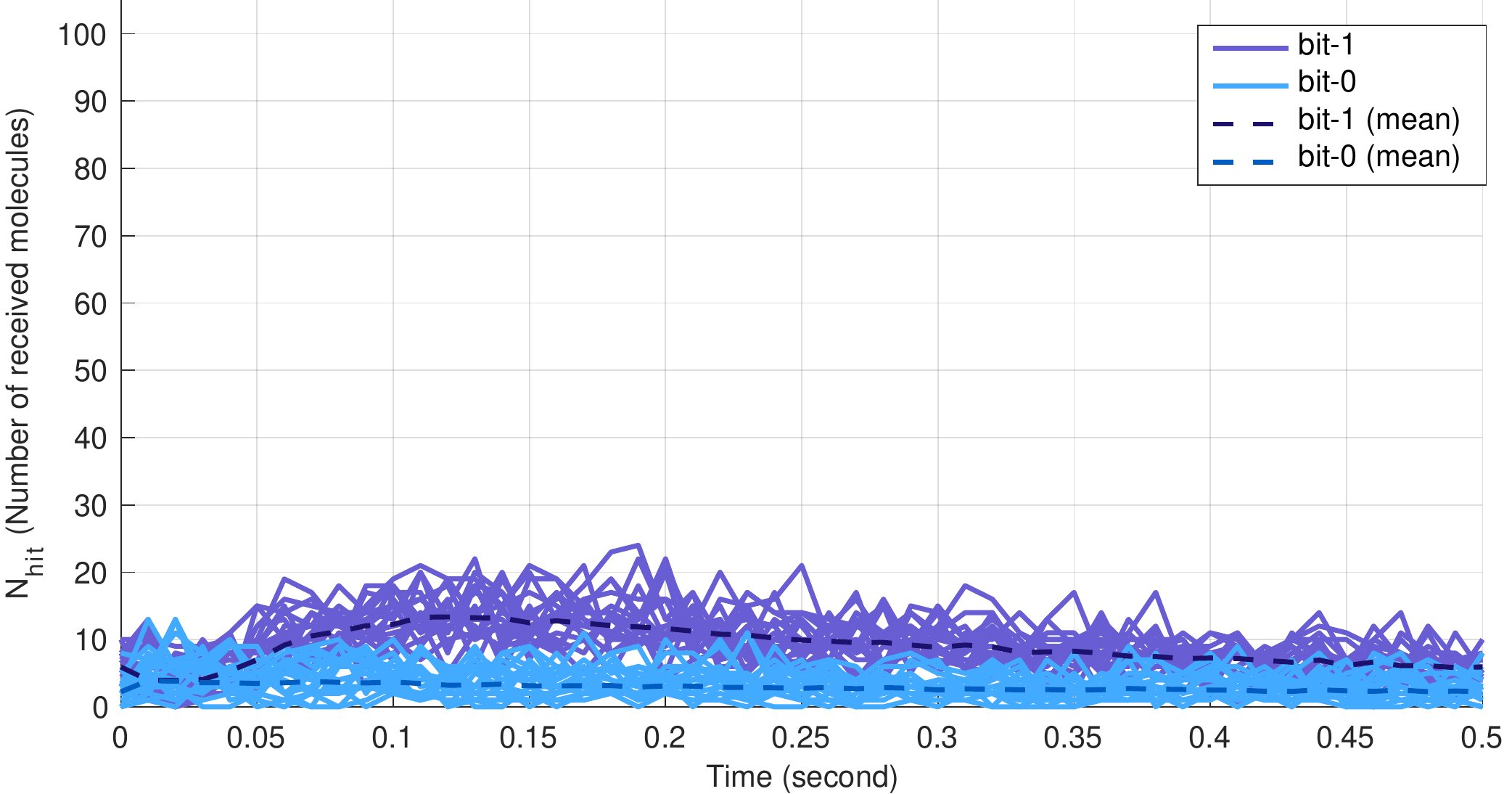}
        \label{fig_eye_diagram_bad_withPA}}
	\end{center}
   \caption{MOL-Eye diagrams of consecutive bit transmissions in different environmental conditions for both with and without CPA cases ($\symdur=\SI{0.5}{\second}$).} 
   \label{fig_eye_diagrams}
\end{figure*}
\subsection{Eye Diagram Analysis}
In the context of MC, we define three new metrics, which are standard deviation of the number of received molecules, $\mystd{c_0(:)}$ and $\mystd{c_1(:)}$, MaxEH, and CSNR. The standard deviation is simply calculated by quantifying the amount of variation in the number of received molecules. MaxEH is the maximum distance between the curves of a bit-0 and \bitone{} in a single symbol slot. In \eqref{eq_integralDif} and \eqref{eq_CSNR}, the steps of CSNR calculation are depicted.

In the eye diagram analysis, we use good, moderate, and harsh environments whose parameters are given in Table~\ref{tbl_env_params}.
\begin{table}[th]
\renewcommand{\arraystretch}{1.2}
\caption{Environment Parameters for The Eye Diagram Analysis}
\label{tbl_env_params}
\centering
  \begin{tabular}{C{2cm} C{1.7cm} C{1.7cm} C{1.7cm}}
    \hline
    Environments & $d (\si{\micro\meter})$ & $D (\si{\micro\meter^2/\second})$ & $\vflow (\si{\micro\meter/\second})$ \\ \hline
    Good  & 4 & 150 & 5 \\
    Moderate & 5 & 100 & 2.5 \\
    Harsh & 6 & 50 & 0 \\
    \hline
  \end{tabular}
\end{table}

Table~\ref{metrics_eye} shows the $\mystd{c_0(:)}$, $\mystd{c_1(:)}$, MaxEH, and CSNR values in these three environments. As seen in the table, MaxEH and CSNR increase while the standard deviation decreases as the environment gets better or when BCSK-CPA method is used. For the calculation of MaxEH, we normalize the total number of received molecules. For ease of comparison, the metric values of BCSK-CPA in the good environment are given in a bold face font, which represent the best results among six different conditions.
\begin{table}[th]
\renewcommand{\arraystretch}{1.2}
\caption{Metrics of Eye Diagram}
\label{metrics_eye}
\centering
  \begin{tabular}{C{2cm} C{1.7cm} R{1.7cm} R{1.7cm}}
    \hline
    Environments & Metric Name & with CPA & w/o CPA \\ \hline
    \multirow{4}{*}{Good}  & $\mystd{c_0(:)}$ & \textbf{11.0948} & 11.5592 \\
          & $\mystd{c_1(:)}$ & \textbf{29.3192} & 29.8338 \\    
          & MaxEH & \textbf{127.6994} & 118.0000 \\    
          & CSNR & \textbf{14.5762} & 11.6322 \\    \hline
 \multirow{4}{*}{Moderate} & $\mystd{c_0(:)}$ & 13.8048 & 15.1311 \\
          & $\mystd{c_1(:)}$ & 27.2424 & 29.1996 \\    
          & MaxEH & 68.0454 & 65.0000 \\    
          & CSNR & 8.5072 & 6.6060 \\    \hline
    \multirow{4}{*}{Harsh} & $\mystd{c_0(:)}$ & 16.0739 & 19.7512 \\
          & $\mystd{c_1(:)}$ & 22.7202 & 27.6683 \\    
          & MaxEH & 38.3462 & 36.0000 \\    
          & CSNR & 3.6683 & 2.8110 \\   
    \hline
  \end{tabular}
\end{table}

Fig.~\ref{fig_flow_vs_CSNR} depicts the effect of flow velocities over $\diCO$ values and modulation techniques. As seen in this figure, unlike the relation between BER and flow velocity as in Fig.~\ref{fig_validation_flow_vs_BER}, CSNR rises with increasing flow velocity as expected.

Finally, the eye diagrams for three different environmental conditions can be seen in Fig.~\ref{fig_eye_diagrams}. The widths are wider in the good environment compared to the moderate and the harsh environments, which shows that the effect of ISI is less in good environment. Moreover, the eyes are also more open in the good environment than the moderate and harsh the environments, which shows that there is less noise in the good environment. Please note that the received signals in Fig.~\ref{fig_eye_diagrams} are obtained from consecutive transmissions and they include ISI.

\section{Conclusion and Future Work}
In this paper, we have proposed a new metric called MOL-Eye based on the conventional eye diagram concept. We introduced three new metrics for performance evaluation using derivatives of MOL-Eye i.e., i) MaxEH, standard deviation of the number of received molecules, and CSNR. We showed that these metrics can be used to exhibit the quality of different performance enhancement methods in MC (i.e., BCSK vs. BCSK-CPA). In our experiments, we considered a vessel like environment with laminar flow under three different environmental conditions (i.e., good, moderate, harsh) for two different modulation techniques, namely conventional BCSK and BCSK-CPA (which was also proposed in this paper as an alternative power adjustment method to BCSK-PA). When we compared the performances under different conditions, we confirmed that BCSK-CPA outperforms BCSK and the good environment case outperforms the moderate and harsh cases, and so on as expected. Based on our evaluations, we have also seen that CSNR is inversely proportional with BER as in Fig.~\ref{fig_CSNR_vs_BER}. Moreover, CSNR and BER have one-to-one relation, which points out that BER can be formulated in terms of CSNR. As a future work, we plan to find the analytical derivation of the relation between CSNR and BER.





\bibliographystyle{IEEEtran}
\bibliography{Eye_Diagram_Enh_2017}

\end{document}